\documentclass[pdflatex,sn-mathphys-num]{sn-jnl}

\usepackage{graphicx}%
\usepackage{multirow}%
\usepackage{amsmath,amssymb,amsfonts}%
\usepackage{amsthm}%
\usepackage{mathrsfs}%
\usepackage[title]{appendix}%
\usepackage{xcolor}%
\usepackage{textcomp}%
\usepackage{manyfoot}%
\usepackage{booktabs}%
\usepackage{algorithm}%
\usepackage{algorithmicx}%
\usepackage{algpseudocode}%
\usepackage{listings}%
\usepackage{lineno}

\theoremstyle{thmstyleone}%
\theoremstyle{thmstyletwo}%

\theoremstyle{thmstylethree}%

\newcommand{\RevA}[1]{{\color{black}#1}}

\begin{document}

\title[HydroX]{HydroX, a light dark matter search with hydrogen-doped liquid xenon time projection chambers}


\author*[1]{\fnm{W.H.} \sur{Lippincott}}\email{hlippincott@ucsb.edu}

\author[1]{\fnm{H.N.} \sur{Nelson}}

\author[2,3]{\fnm{D.S.}\sur{Akerib}}
\author[1]{\fnm{C.}\sur{Amarasinghe}}
\author[2,3]{\fnm{A.}\sur{Ames}}
\author[4]{\fnm{H.M.}\sur{Ara\'{u}jo}}

\author[1]{\fnm{J.W.}\sur{Bargemann}}
\author[5]{M.C.~Carmona-Benitez}

\author[2,3]{\fnm{R.}\sur{Coronel}}
\author[6,7]{\fnm{C.E.}\sur{Dahl}}
\author[8]{\fnm{S.}\sur{Dey}}

\author[5]{J.~Genovesi}
\author[9]{\fnm{S.J.}\sur{Haselschwardt}}
\author[4]{\fnm{E.}\sur{Jacquet}}
\author[10]{\fnm{D.}\sur{Khaitan}}
\author[11]{\fnm{D.}\sur{Kodroff}}
\author[12]{\fnm{S.}\sur{Kravitz}}
\author[9]{\fnm{W.}\sur{Lorenzon}}
\author[2,3]{\fnm{S.}\sur{Luitz}}
\author[11]{\fnm{A.}\sur{Manalaysay}}
\author[13]{\fnm{C.}\sur{Maupin}}
\author[2,3,14]{\fnm{M.E.}\sur{Monzani}}
\author[4]{\fnm{K.C.}\sur{Oliver-Mallory}}
\author[11]{\fnm{E.}\sur{Perry}}
\author[10]{\fnm{Y.}\sur{Qie}}
\author[2,3]{\fnm{T.}\sur{Shutt}}
\author[7]{\fnm{D.J.}\sur{Temples}}
\author[1]{\fnm{M.}\sur{Trask}}
\author[2,3]{\fnm{A.}\sur{Wang}}
\author[10]{\fnm{F.L.H.}\sur{Wolfs}}
\author[11]{\fnm{D.}\sur{Woodward}}
\author[1]{\fnm{R.}\sur{Zhang}}
\author[11,15]{\fnm{T.}\sur{Zhang}}

\affil*[1]{University of California, Santa Barbara, Department of Physics, \orgaddress{\city{Santa Barbara}, \state{CA} \postcode{93106-9530}, \country{USA}}}
\affil[2]{\orgname{SLAC National Accelerator Laboratory}, \orgaddress{\city{Menlo Park}, \state{CA} \postcode{94025-7015}, \country{USA}}}
\affil[3]{\orgdiv{Kavli Institute for Particle Astrophysics and Cosmology}, \orgname{Stanford University}, \orgaddress{\city{Stanford}, \state{CA} \postcode{94305-4085}, \country{USA}}}

\affil[4]{Imperial College London, Physics Department, Blackett Laboratory, London SW7 2AZ, UK}

\affil[5]{Pennsylvania State University, Department of Physics, University Park, PA 16802-6300, USA}

\affil[6]{Northwestern University, Department of Physics \& Astronomy, Evanston, IL 60208-3112, USA}
\affil[7]{Fermi National Accelerator Laboratory (FNAL), Batavia, IL 60510-5011, USA}

\affil[8]{University of Oxford, Department of Physics, Oxford OX1 3RH, UK}

\affil[9]{University of Michigan, Randall Laboratory of Physics, Ann Arbor, MI 48109-1040, USA}
\affil[10]{University of Rochester, Department of Physics and Astronomy, Rochester, NY 14627-0171, USA}
\affil[11]{Lawrence Berkeley National Laboratory (LBNL), Berkeley, CA 94720-8099, USA}

\affil[12]{University of Texas at Austin, Department of Physics, Austin, TX 78712-1192, USA}

\affil[13]{South Dakota Science and Technology Authority (SDSTA), Sanford Underground Research Facility, Lead, SD 57754-1700, USA}
\affil[14]{Vatican Observatory, Castel Gandolfo, V-00120, Vatican City State}
\affil[15]{University of California, Berkeley, Department of Physics, Berkeley, CA 94720-7300, USA}


\abstract{Experimental efforts searching for dark matter particles over the last few decades have ruled out many candidates led by the new generation of tonne-scale liquid xenon. For light dark matter, hydrogen could be a better target than xenon as it would offer a better kinematic match to the low mass particles. This article describes the HydroX concept, an idea to expand the dark matter sensitivity reach of large liquid xenon detectors by adding hydrogen to the liquid xenon. We discuss the nature of signal generation in liquid xenon to argue that the signal produced at the interaction site by a dark matter-hydrogen interaction could be significantly enhanced over the same interaction on xenon, increasing the sensitivity to the lightest particles. We discuss the technical implications of adding hydrogen to a xenon detector, as well as some background considerations. Finally, we make projections as to the potential sensitivity of a HydroX implementation and discuss next steps.  }




\maketitle

\section{Introduction}\label{sec1}

Astrophysical measurements across all scales from the smallest dwarf galaxies to the cosmic microwave background agree that most of the matter in our universe is made up of dark matter (DM)~\cite{bertone:2004pz,Feng:2010gw}.  None of the particles in the Standard Model of particle physics can account for the DM, and the observation of a DM particle would be a revolutionary breakthrough and evidence for new physics.  A leading hypothesis for DM is a thermal relic from the Big Bang with a mass between about 1 GeV and 100 TeV,
with several candidates predicted by extensions to the Standard Model, often called Weakly Interacting Massive Particles (WIMPs)~\cite{jungman:1995df,Feng:2010gw,bertone:2004pz}. Candidates can interact via both spin independent (SI) and spin dependent (SD) elastic scattering, along with a wider set of more complex interactions that are also important channels for detection~\cite{fitzpatrick:2013lia, Gluscevic:2015sqa}. 
To date, no WIMPs have been observed, with a series of experiments setting increasingly stringent limits on the cross section for DM scattering with normal matter. The current best constraint on DM with masses above 10 GeV/c$^2$ comes from the LZ experiment, a large liquid xenon time projection chamber (LXe-TPC); two other LXe-TPCs, XENONnT and PandaX-4T have also published strong constraints~\cite{LZ:2022lsv,XENON:2023cxc,PandaX-4T:2021bab}. 

A largely unexplored window for thermal dark matter from 10 GeV down to MeV mass scales can be opened up by allowing new force mediators into the theory,  below which point constraints from structure formation in the universe take effect.  
Extensions to the Standard Model produce several electroweak candidates below 1 GeV that evade all constraints from the LHC~\cite{Kozaczuk:2013spa,Cerdeno:2011qv}. Other frameworks naturally produce candidates at $\sim$1 GeV, as in Asymmetric Dark Matter~\cite{Lin:2011gj}. 
A summary of the case for low mass dark matter can be found in the 2017 report from the US Cosmic Visions DM working group~\cite{Battaglieri:2017aum}, the 2019 DOE New Initiatives in DM Report~\cite{BRN:2019}, and the white papers produced for Snowmass 2021~\cite{Cooley:2022ufh,Akerib:2022ort,Essig:2022dfa}. Over the past several years, several experiments have begun probing this region of parameter space via both DM-nucleus and DM-electron scattering, for example NEWS-G, SENSEI, DAMIC, and SuperCDMS-HVeV as well as XENON, LZ, and DarkSide-50 through the Migdal effect~\cite{NEWS-G:2017pxg,Aguilar-Arevalo:2019wdi,Abramoff:2019dfb,agnese:2018col,LZ:2023poo, XENON:2019zpr, DarkSide:2022dhx}. As for the higher masses, no observations have yet been reported.


For kinematic reasons, low mass DM particles do not transfer energy to heavy nuclei like xenon as efficiently as heavier WIMPs. The smallest DM mass $m_\chi$ that a detector with target atomic mass $A$ and nuclear recoil energy threshold $E_\mathrm{NR}$ can detect is approximately~\cite{Battaglieri:2017aum}:
\begin{equation}
   m_{\chi}=88\,\mathrm{MeV}\times\left[\frac{E_{\mathrm{NR}}}{0.1\,\mathrm{keV}}\times A\right]^{1/2}.
   \label{eq:lowmass}
\end{equation}
For xenon recoils in LXe TPCs, $E_{\mathrm{NR}}\!\sim\!5\,$keV for standard conditions, giving sensitivity to $\sim$8 GeV DM particles. 




In this article we discuss HydroX, an idea to deploy hydrogen, the nucleus with the smallest $A$,
into a LXe detector at the level of one percent by number density, providing $\sim1\,$kg in a detector the size of LZ.  The proton that constitutes the hydrogen nucleus becomes the dark matter target, with LXe as the sensor. The use of D$_2$ instead of H$_2$ can bring neutrons into the interaction. Thus, HydroX can provide sensitivity to both SI and SD channels for both DM-proton and DM-neutron interactions.  No other set of target nuclei provides sensitivity to this broad menu of interaction types for DM masses below the proton mass, and,
as many DM models predict suppressed leptonic interactions, there is a clear need for experiments designed to detect all possible DM-nucleus interactions at these mass scales. 



\section{Methods}\label{sec2}
\subsection{LZ as a Case Study}\label{sec:LZ}
To study the potential of HydroX, we start by assuming it can be carried out in the LZ detector as described in~\cite{Akerib:2019fml}, but most of the arguments can be carried to other detector implementations or a next generation LXe-TPC~\cite{Aalbers:2022dzr, XLZD:2024nsu}.
LZ consists of a two-phase LXe-TPC containing 7 tonnes of fully active LXe. Xenon is a liquid around 165 K at 1 atmosphere, and the entire inner detector is a cylindrical, cryogenic volume. In a LXe-TPC, particle interactions can create prompt scintillation light, known as S1, and electron-ion pairs. Freed electrons are drifted through the liquid by an electric field perpendicular to the liquid-gas interface, extracted into the gas region, and then accelerated through a stronger electric field, creating proportional scintillation light, known as S2, that is a direct measure of the number of extracted electrons. Photomultiplier tubes (PMTs) collect both S1 and S2 light, with $xy$ position reconstruction provided by the PMT hit pattern of the S2 light, and $z$ reconstruction by the timing between S1 and S2. 


The main objective in designing a dark matter detector is elimination of background radiation that can mimic a dark matter signal.  LZ has been painstakingly constructed out of low radioactivity components in a clean environment to minimize radioactive contamination~\cite{LZ:2020fty}. Because of the high density of LXe ($\sim$3 g/cm$^3$ at 165~K), external radioactivity is attenuated before it can reach an innermost, ``fiducial'' volume at the core of a large LXe volume. To further reduce backgrounds, the LZ detector includes two active veto regions outside the main TPC to  tag particles generated in detector materials that  scatter in the TPC before exiting. 
LZ and its veto detectors sit in a water tank for further shielding, located in the 4,850-foot level of the Sanford Underground Research Facility, in Lead, South Dakota, USA. 

\subsection{Modeling DM-proton scattering in LXe}

\begin{figure}
    \centering
\includegraphics[width=0.8\textwidth]{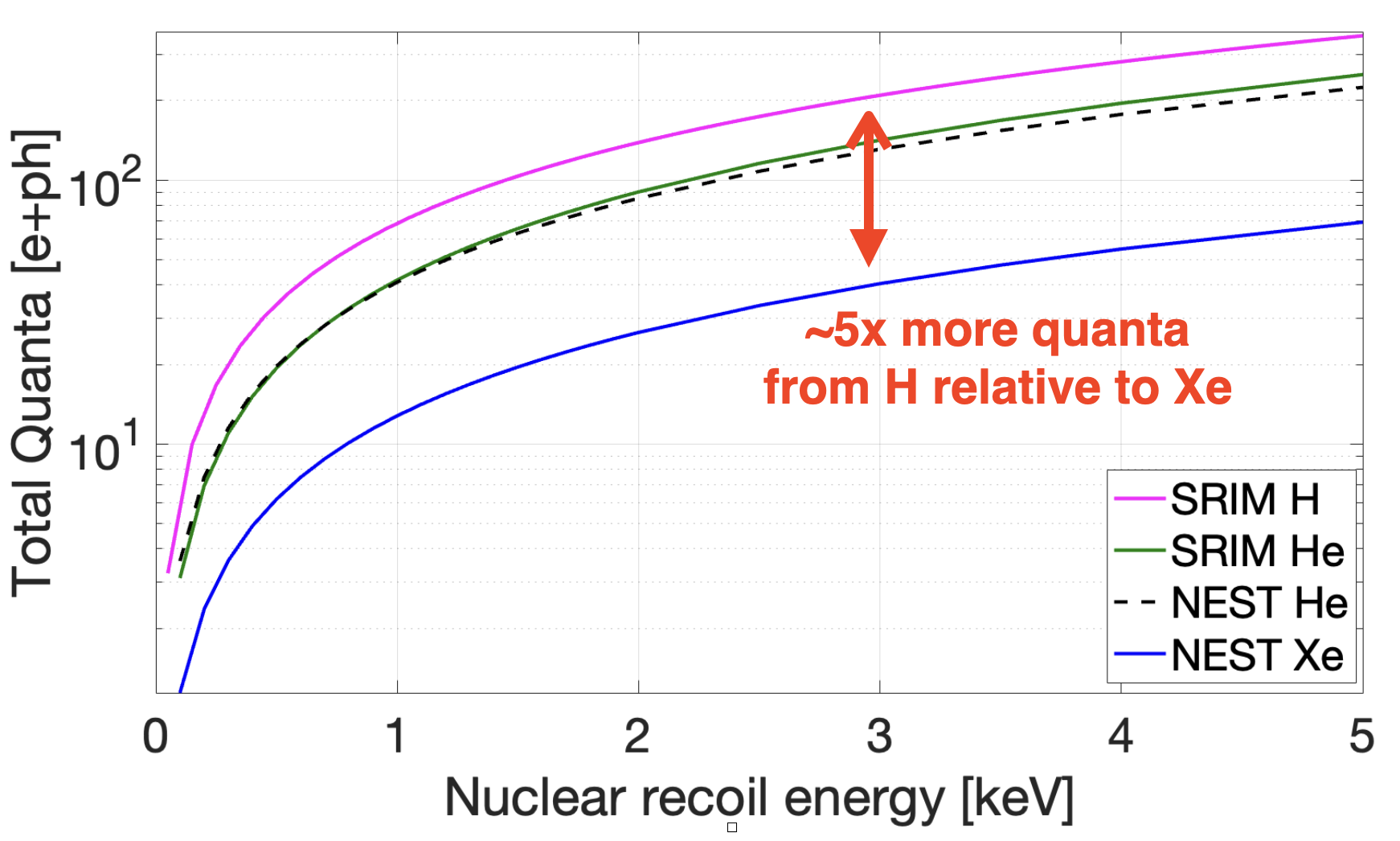}
\caption{\label{fig:Lind} Total quanta (photons and electrons) released from various recoils in LXe. The blue curve shows the quanta for low energy Xe recoils in LXe as modeled by NEST. The magenta line shows a SRIM calculation~\cite{Ziegler:2010a} for H recoils, assuming only electronic stopping contributes, with about 5$\times$ more quanta produced per unit energy relative to Xe. The green and dashed black lines show two independent predictions for yields from He recoils in LXe, one from SRIM package and the second from NEST. The two models for helium are in good agreement. }
\end{figure}

 For sensitivity to galactic DM particles
with mass less than 1 GeV, kinematics dictate that the target particle must be light
and/or the sensor must have a low energy threshold
(see Eq.~\ref{eq:lowmass}). As the lightest element, hydrogen being a single proton is kinematically the most attractive nucleus for the detection of nuclear recoils from sub-GeV DM particles that interact with quarks and/or gluons (the molecular binding energy of H$_2$ is 4.75 eV, and single proton recoils will be the signal for DM scattering). Less obvious is the potential very low 
energy threshold in LXe
for the detection of a proton 
recoil.  In a LXe-TPC, most of the energy from electronic recoils (ER) is transferred to the electronic system of the medium, producing electronic excitation and ionization (electronic stopping) that lead to S1 and S2. For nuclear recoils (NR), the recoiling atom primarily experiences quasi-elastic collisions with other xenon atoms mediated by the interatomic potential (nuclear stopping). Energy transferred in these elastic collisions does not produce S1 and S2 signals and is eventually lost to heat, but an inelastic transfer of energy through slow disruption of the xenon electron cloud can occur, producing S1 and S2 with reduced yield and a different S1/S2 ratio compared to ER.
Measurements of nuclear recoil yields in LXe have shown that 
for energies below 10~keV, less than $\sim$20\% of the 
total xenon recoil energy ends up as visible quanta (photons and electrons). The Noble Element Simulation Technique (NEST) is a widely-used, data-driven simulation package that models electron and photon yields in LXe~\cite{szydagis:2011tk,szydagis:2013sih,szydagis_m_2018_1314669}, and the yellow curve in Fig.~\ref{fig:Lind} shows the total number of quanta released for low energy xenon recoils in LXe as modeled by NEST.

In contrast, a recoiling proton will transfer very little of its energy via elastic collision to xenon atoms, similar to the case of
a fast neutron incident on a high-$A$ target, as described
by Fermi~\cite{fermi:1946} or Segre~\cite{Segre:1977}.
In the limit of hard-sphere proton-xenon collisions, on average only $\sim$1.5\% of the proton kinetic energy is lost to xenon kinetic energy per elastic collision; actual proton-xenon collisions are softer
than hard-sphere, reducing the average energy loss~\cite{Krstic:2006}.  
A typical
proton struck by a proton-mass galactic dark
matter particle will survive for at least 100 subsequent elastic proton-xenon collisions
prior to falling below the xenon ionization
threshold energy (in the
hard-sphere limit) with each collision providing an opportunity for  excitation of the electrons bound in the xenon atom.  
By contrast, a typical xenon atom struck by a $100\,$GeV galactic dark matter particle will fall beneath
the threshold  for xenon ionization after an average of only about 8 collisions.
Because a proton transfers less of its kinetic energy
to translation of the sensor xenon atoms, more of the
 kinetic energy is available for electronic
excitation and ionization, leading to more charge and light production.
We estimate the raw electronic and nuclear stopping powers for hydrogen recoils in LXe using the Stopping and Range of Ions in Matter (SRIM) simulation package~\cite{Ziegler:1985,Ziegler:2010a}, and the predicted signal yield is shown by the magenta curve in Fig.~\ref{fig:Lind} under the assumption that all of the electronic stopping goes into signal but none of the nuclear stopping does. We estimate that H recoils in LXe will result in about 5 times more quanta released per unit energy relative to Xe recoils.

There are no data measuring yields for H recoils in LXe; there exist measurements for alpha recoils at high energy, for which a similar signal enhancement can be expected~\cite{szydagis_m_2018_1314669,dahl:2009nta,bradley:2014lga}. \RevA{While the alpha data are taken at MeV-scale energies, the NEST package includes a power-law extrapolation down to low energies~\cite{rischbieter2018cpad}, shown by the dashed black line in Fig.~\ref{fig:Lind}.} Our SRIM calculation for a He projectile is shown in green, showing good agreement with the NEST extrapolation, particularly given the different inputs and assumptions made by the two models. This agreement combined with the physical intuition already described gives some confidence in the expectation that proton recoils will produce significantly more photons and electrons than Xe recoils. 

A second key question is how the increased signal is partitioned into S1 and S2 pulses for proton recoils.  As the S2:S1 ratio provides event-by-event discrimination between ER and NR signals in a LXe-TPC, it will be crucial to understand whether ER backgrounds will be distinguishable from proton recoils. \RevA{The number of ionization electrons in an event is determined by a complex interaction of the drift field, the electron thermalization length, and the track density. Electrons and ions can recombine, reducing the number of electrons in favor of scintillation photons. The presence of H$_2$ might speed up the thermalization process, in principle leading to more recombination by increasing the density of the electrons and ions. }

 At low energies, where electron-ion recombination is less important, the S2:S1 ratio is determined by the partition of electronic stopping into initial ionization and excitation of xenon atoms. The kinematics
of electronic excitation by a heavy projectile depend primarily on its velocity with respect to the
target~\cite{Fermi:1947uv}. 
A proton with kinetic energy of $1\,$keV has the same velocity as a xenon atom with kinetic energy
$(m_{\mathrm{Xe}}/m_{\mathrm{p}})\!\times\!1\,\mathrm{keV}=130\,$keV. 

A recent measurement of the response of LXe to degraded alpha particles, or helium recoils, showed a partitioning that was quite xenon-like with excellent discrimination against electronic recoils~\cite{Haselschwardt:2023iqn}. 
As it is not fully understood what drives the partitioning between excitation and ionization for xenon recoils in LXe, it is hard to predict exactly what S2/S1 distribution hydrogen recoils might produce, but it is reasonable to take the electron recoil and xenon recoil distributions as bounding cases. 

\subsection{Hydrogen-doped LXe-TPC performance}
\subsubsection{Possible S1 and S2 signal losses to hydrogen}
The presence of molecular species in LXe can reduce the amount of both S1 and S2 signal collected via quenching or absorption of light and charge. H$\mathbf{_2}$ is not an electronegative species and should not absorb drifting electrons; the MuCap experiment operated a high pressure H$_2$ gas TPC at a similar H$_2$ density ($\sim$0.7 kg/m$^3$) as we hope to achieve in HydroX without seeing significant electron loss~\cite{Egger:2014bua}. Similarly, H$_2$ has no absorption features near 175 nm~\cite{Fujii:2015qqq}, the wavelength of LXe scintillation light, as the Lyman-$\alpha$ series and similar excited molecular states end at $\sim$122 nm, and the Ballmer series picks up above 360 nm~\cite{Herzberg:1950,Keller:2013a,Shaw:2005xm}. 

We expect both the S1 and S2 signals to be quenched by the presence of hydrogen, but for different reasons. LXe scintillation light is produced in the decay of metastable Xe$_2^*$ molecules formed by the combination of excited and ground state xenon atoms, Xe$^*$ + Xe. The excited states, both atomic and molecular, can transfer their energy to H$_2$ molecules via collision, quenching the excitations before a photon is emitted, an effect observed in various combinations of GXe, LAr, and impurities like N$_2$, CO$_2$ and CH$_4$~\cite{Acciarri:2008kv,Pushkin:2006a,Henriques:2017rlj}.

The presence of a significant concentration of molecular H$_2$ in the gas phase will also limit electro-luminescent amplification through a similar competition with energy transfer into molecular excitations. A  simulation of the mixture of gases in the Garfield simulation package~\cite{veenhof:1993hz} suggests that adequate S2 production can be preserved for gas mixtures of over 5\% percent by mole in the gas phase, and we expect that some gain can be recovered by operating at a higher amplification voltage than in the xenon-only run.  A nominal operating condition for H$_2$ doped in LZ would be a field of 10 kV/cm in the gas phase, 25$\%$ above the current LZ operating condition, producing approximately 74 photons per electron. We generally assume that single extracted electrons will still produce enough signal to be readily detected. 

Both primary scintillation and ionization signals for H$\mathrm{_2}$ mole fractions up to $\mathrm{5.7\,}$\% have been observed in a 26~atm gaseous xenon TPC
for the signal from $^{241}$Am $5.5\,$MeV alpha particles~\cite{Tezuka:2004gza}, with 50\% losses in both channels reported for an H$_2$ mole fraction of $1.1\,$\%.  As dual phase LXe-TPCs require the achievement of impurity levels about 20 times more stringent than achieved in~\cite{Tezuka:2004gza}, 
both of the reported losses might be reduced at higher purity. The amount of H$_2$ that can be practically dissolved in the LXe is addressed further in the "Dissolving hydrogen in LXe" subsection of Methods.


An accurate calibration of S1 yields vs liquid H$_2$ concentration may enable a precise \textit{in situ} measurement of H$_2$ loading in a LXe-TPC with standard internal calibration sources, a silver lining that would partly offset the loss of S1 signal. 

Finally, it is worth pondering on how electron transport properties are modified by doping, as this may affect the primary and additional science cases of these experiments. Estimates using \emph{ab initio} calculations for molecular solutes in liquid xenon~\cite{Boyle_2024} confirm the cooling of drifting electrons relative to the pure liquid: the addition of molecular hydrogen increases the drift velocity by a factor of $\sim$2 for percent-level doping, while reducing the longitudinal diffusion constant also by a factor of $\sim$2~\cite{BoylePrivate}. Together, these enable sharper S2 pulses, leading to better clustering of small S2 signals and improved multiple-scatter resolution. The latter is an important parameter to discriminate $\gamma$-ray backgrounds in neutrinoless double-beta decay searches, which will be prototyped in LZ and constitute a key science case for future searches such as XLZD~\cite{XLZD:2024pdv}. The faster electron drift is itself useful as it reduces accidental coincidence backgrounds, as discussed later.
\subsubsection{Technical Feasibility}

A LXe-TPC should maintain high voltage and light collection performance in the presence of H$\mathbf{_2}$. The S1 and S2 light signals in H$_2$-doped LXe will be at the same wavelength as in pure LXe, because energy will transfer from recoiling protons to xenon atoms and electrons. Over longer time scales, it is well known that PMTs can be damaged by exposure to helium and other light gases,
which can diffuse through the PMT glass and cause afterpulsing (e.g. in~\cite{Incandela:1987dh}).
The permeabilities of He, H$_2$, and D$_2$ in fused silica glass have been well measured at temperatures above room temperature~\cite{norton:1953, altemose:1961, Swets:1961, Lee:1962, Lee:1963, shelby:1977, Williams:1922}, and 
 H$_2$ (and D$_2$) mobility is slower relative to helium because of its 10$\%$ larger kinetic diameter~\cite{Mehio:2014}. Mechanical diffusion of all species is exponentially suppressed with decreasing temperature according to the Arrhenius equation,
\begin{equation}
    K = K_0\,\mathrm{exp}(-\Delta H/RT),
    \label{eq:Arrhenius}
\end{equation}
as less thermal energy is available to open pathways (one reason why very few measurements exist below 0$^\circ$C). Here, $K$ is the permeability, $K_0$ is a reference permeability at  a particular temperature, $\Delta H$ is the enthalpy of activation of the glass, $R$ is the ideal gas constant, and $T$ is the temperature. 
Based on the existing data extrapolated to lower temperatures, exposure to an atmosphere of H$_2$ at liquid xenon temperatures does not pose a significant risk to PMTs for run durations small compared to 80~years.

The cryogenic implications of hydrogen doping will be the major technical challenge for 
any implementation of HydroX. LZ has an extensive xenon handling system  to circulate xenon at 400 slpm through a getter to meet the strict electronegative impurity requirements of LXe-TPCs, and
the existing circulation geometry does not easily allow for the presence of a non-condensible gas in the system~\cite{Jensen:1988}.  To mitigate this problem, our initial approach is to  introduce additional stages to the circulation system to remove H$_2$ from the process before the xenon enters the purifier and then re-inject H$_2$ into the liquefied xenon as it returns to the detector. Possible methods for separation include distillation, sparging, or membrane separation~\cite{Adhikari:2006,Li:2018}. For example,  based on preliminary calculations with the McCabe-Thiele method~\cite{Mccabe:2005}, a distillation column with about 50 stages and a reflux ratio of 6.5 would be able to purify the xenon to a level where it could be recirculated for a year while building up less than 1 g of H$_2$ in the existing LZ getter. In the field, the DarkSide-50 experiment operated a 58 stage column to purify underground argon at Fermilab~\cite{Agnes:2018fwg, Back:2012tx}, while XENON has operated an online distillation column to remove Rn, Ar, and Kr impurities~\cite{XENON100:2017gsw,XENON:2021fkt}.  The extracted H$_2$ gas would then be re-injected into the liquid xenon using a membrane as the liquid re-enters the central TPC, maintaining a constant H$_2$ partial pressure head above in the gas phase above. Significant R\&D on the circulation system will be required before implementation on a system of sufficient scale to be representative of a large detector.

\subsection{Dissolving hydrogen in LXe}
\label{sec:Henry}

\begin{figure} 
    \centering
\includegraphics[width=0.8\textwidth]{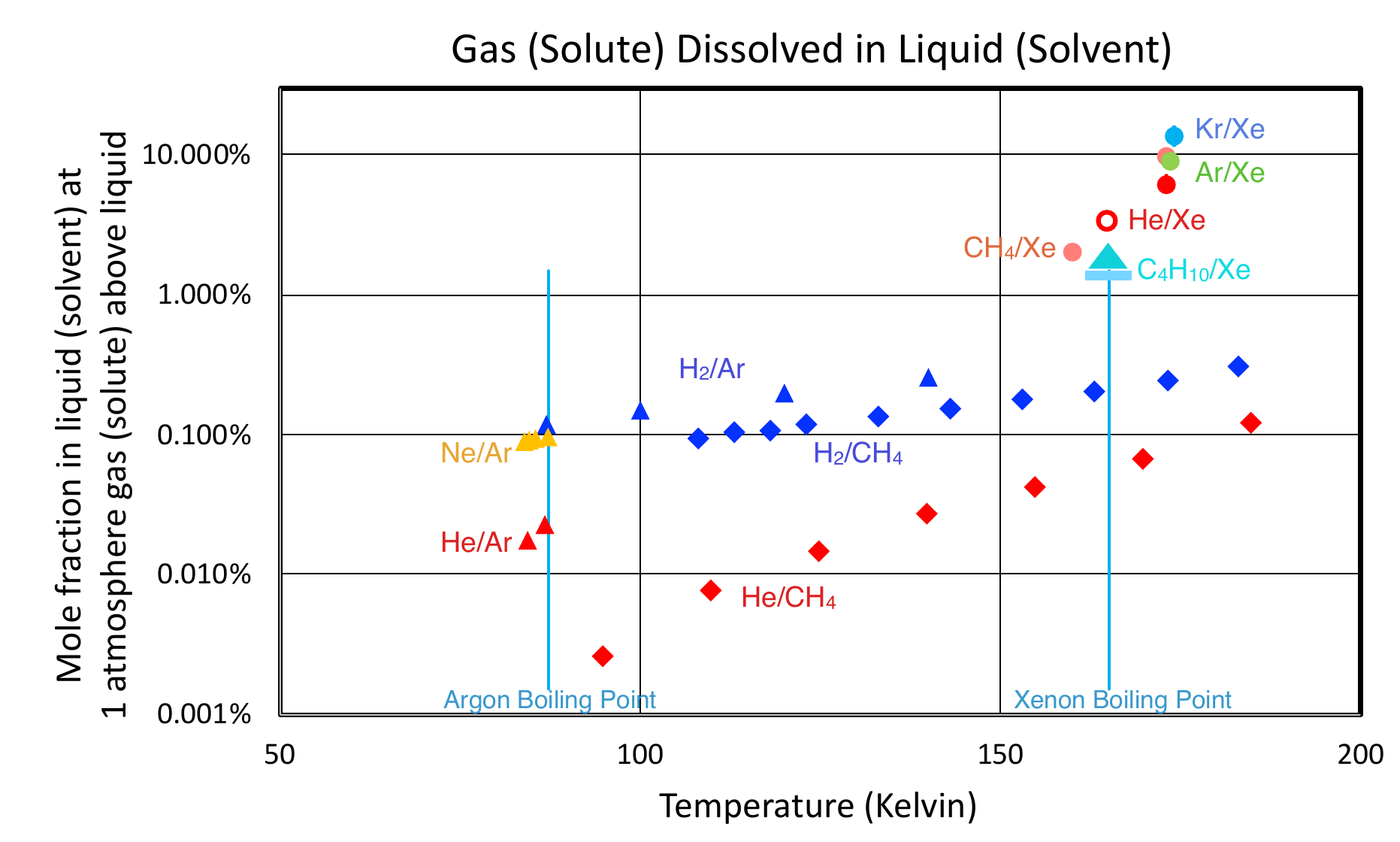}
\caption{\label{fig:ManyEquilibrium}Solubilities of various gases in cold liquids as a function of temperature. Methane (argon/xenon) as a solvent are shown in diamonds (triangles/circles).  The various solutes are labeled and distinguished by color. The open red circle shows unpublished data from a test stand at Fermilab. The solubilities in liquid xenon measured by LUX and associated small setups are substantially higher than those in methane and argon.}
\end{figure}

 
The sensitivity for HydroX depends directly on the amount of H$_2$ that can be dissolved into the LXe. Small concentrations
 in the LXe mixture are governed by Henry's law: the concentration in liquid is proportional to the vapor pressure of the solute gas introduced above the LXe. We estimate Henry's coefficient based primarily on the scaling of LUX measurements of the concentrations of helium, argon, and krypton in the gas and liquid phases of xenon at $173\,$K for a xenon
 vapor pressure of $1.56\,$atm.  
 
Figure~\ref{fig:ManyEquilibrium} shows the solubilities of several  mixtures~\cite{Karasz:1958,Volk:1960,Heck:1967,Shukla:1986,Yoshino:1976}, including unpublished measurements from LUX and at a test stand at Fermilab. 
The test stand measurements give a factor of two smaller liquid concentration than the LUX value, a difference that results in part from different temperature conditions.  
Xenon is substantially more efficient as a solvent than other noble gases, a consequence of the deeper Van der Waals potential~\cite{Pollack:1991,Rentzepis:1981}. In particular, the equilibrium concentration of
 H$_2$ (or D$_2$) is likely to exceed that of helium: increased solubility of a gas in a liquid is strongly correlated with a larger minimum in the Van der Waals potential~\cite{PLG:1986}, and the depth of the minimum in the H$_2$-Xe potential of $8.0\,$meV~\cite{LeRoy:1987} is more than three times that of the He-Xe potential~\cite{Danielson:1988}. While these considerations suggest that HydroX can achieve adequate H$_2$ loading, no measurements of H$_2$ solubility in LXe currently exist, and a careful characterization of the solubilities
 in LXe of H$_2$ will be required to understand the ultimate sensitivity of HydroX.  

\subsection{Data Availability}
\RevA{The data that support the findings of this study are available from the corresponding author upon reasonable request.}

\section{Results and Discussion}
\subsection{Backgrounds}
By using LZ as the ``host'' detector, the H$_2$ target benefits from the exquisite low background environment of LZ, including the self-shielding provided by a large LXe mass and $xyz$ reconstruction. For self-shielding: consider an external $1\,$MeV $\gamma$-ray entering the detector, which can Compton scatter and cause a background to a WIMP signal.  This $\gamma$-ray has an attenuation length of about $6\,$cm in LXe, which is less than 1/10 of the $75\,$cm radius of LZ. The attenuation length of that same $\gamma$-ray would be 80 times the radius of a 1 m$^3$ high-pressure gas TPC.  The  attenuation length for the same $\gamma$-ray in a hypothetical $2\,$kg liquid hydrogen detector would be 6 times its radius.  

For very low energy depositions like the ones relevant for light WIMP searches, events near detector surfaces can produce significant and difficult to understand backgrounds. For example, the CoGeNT detector observed an increasing spectrum at threshold that has been interpreted as both a dark matter signal but also as a poorly modeled surface background~\cite{aalseth:2014jpa, aalseth:2012if}.  Similar unknown sources of noise are contributing backgrounds to the new wave of ultra-low threshold detectors like SENSEI and SuperCDMS-HVeV
~\cite{Abramoff:2019dfb, agnese:2018col}.
For a hydrogen run in LXe-TPCs, surface effects will be suppressed by the $xyz$ reconstruction, allowing the definition of a central fiducial mass where surface backgrounds are greatly reduced.

Backgrounds can be introduced with H$_2$ or D$_2$, particularly due to tritium contamination; to actually deploy H$_2$ in a low background environment, a tritium relative abundance less than about 1 in $10^{24}$ must be achieved, while atmospheric-derived sources of H$_2$ or D$_2$ have tritium concentrations at the 1 in $10^{18}$ level. \RevA{Two potential solutions to this problem exist. First, distillation can be used to separate tritium from H$_2$. As one example, the MuCap experiment achieved a suppression of deuterium in hydrogen by about 4 orders of magnitude~\cite{Andreev:2012fj,alexseev:2006}; tritium would be more readily suppressed, but work remains to prove this is feasible. A second solution is an underground source of hydrogen, an area of active research for the hydrogen energy industry (e.g.~\cite{H2review}). Solvents used in low-background organic scintillators source some of their materials from underground sources, and underground argon has become a key component in rare event searches~\cite{Alimonti:1998,DarkSide:2015cqb}. We do not know of any explicit measurements of the tritium content from underground sources of hydrogen, but factors of 10-100 suppression do not sound implausible.}  

\RevA{In the sensitivity projections discussed in the next section, we take the background model used in Ref.~\cite{LZ:2018qzl}, with all numbers identical to those in Table III of that reference. We assume tritium can be removed by distillation or underground sources and the hydrogen brings in no other backgrounds. We do not currently assume any accidental backgrounds as observed in recent LXe-TPC detectors; more discussion on this point can be found in the next section.} 




\subsection{WIMP sensitivity}\label{sec:sensitivity}

\begin{figure} 
  \center
\includegraphics[width=0.9\columnwidth]{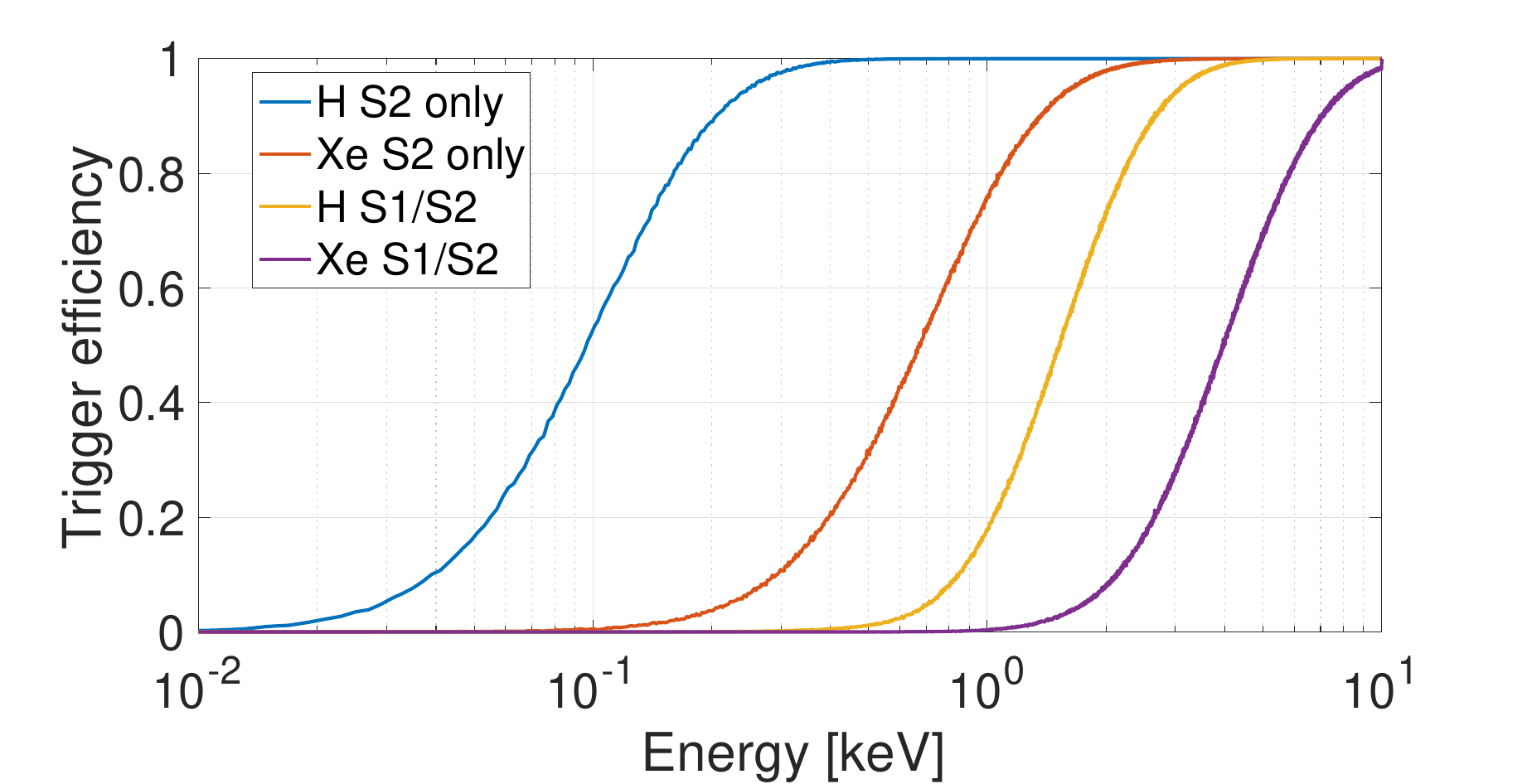}
\caption{\label{fig:threshold} Simulated energy thresholds for Xe and H recoils in LXe. The thresholds are calculated using NEST and the full detector simulation described in~\cite{LZ:2018qzl} and shown for both a standard S1/S2 analysis, requiring an S1 signal in 3 photomultiplier tubes, as well as for an S2-only analysis, requiring 3 detected electrons.}
\end{figure}

To put everything together and calculate the dark matter sensitivity that could be achieved by HydroX, the SRIM hydrogen recoil model is incorporated into NEST and combined with the LZ detector model used in~\cite{LZ:2018qzl} to generate a prediction for the energy threshold for H recoils in a doped LZ detector, shown in Fig.~\ref{fig:threshold}. Based on the high energy alpha data, the NEST model assumes that the S2/S1 splitting is similar to that of electronic recoils. We assume a loss of 50\% of the generated S1 signal to quenching based on~\cite{Tezuka:2004gza} and S2 yields of 240 phd/e following the Garfield model above. We assume 0.95 kg of H$_2$ in the fiducial volume (containing 5.6 tonnes of Xe), for a mole fraction of 1.1\%. The threshold is calculated for a standard S1/S2 analysis, which requires the detection of an S1 signal in at least 3 PMTs as described in~\cite{LZ:2018qzl}. 

\begin{figure} 
  \center
\includegraphics[width=0.6\columnwidth]{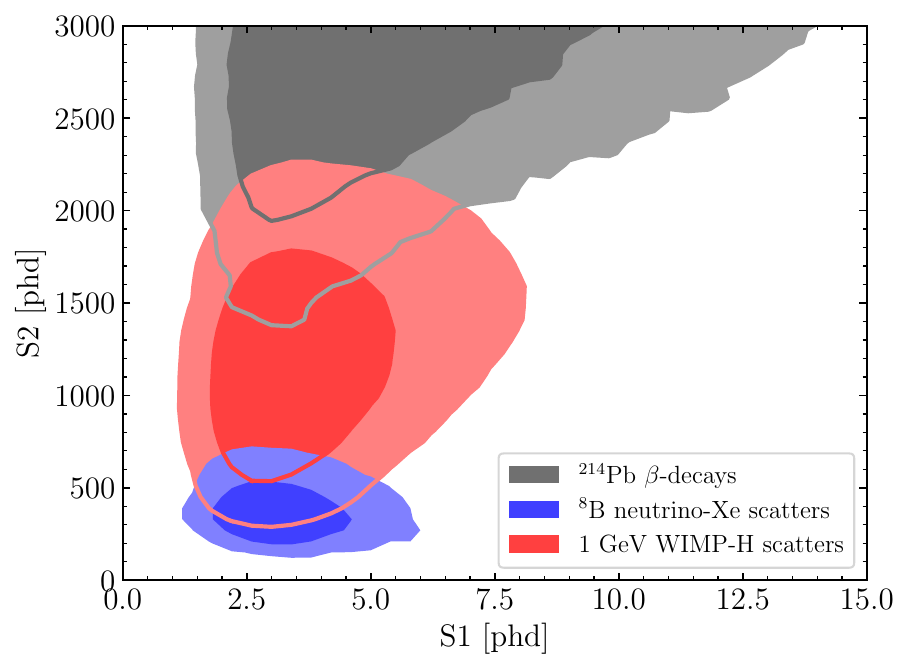}
\caption{\label{fig:S2S1} Simulated contours of S1 and S2 for various event types. The red contours show  1 GeV WIMP-hydrogen scattering, and the blue contours show $^8$B neutrino-xenon scattering. The grey contours show $^{214}$Pb beta decay events. The dark (light) region represent the 1$\sigma$ (2$\sigma$) region.}
\end{figure}

\RevA{Figure~\ref{fig:S2S1} shows simulated S1/S2 distributions for hydrogen scattering from a 1 GeV WIMP, xenon nuclei scattering from $^8$B neutrinos (similar to a $\sim$6 GeV WIMP scattering with xenon), and the $^{214}$Pb beta background. As hydrogen has only one nucleon and no neutrons, its coherent neutrino-nucleus scattering cross section is negligible compared to xenon. The H-recoil spectrum falls below the beta background because it is primarily populated by events with upward fluctuations in S1 and  downward fluctuations in S2 (a similar argument applies for the $^8$B spectrum compared to a xenon recoil band or a higher energy WIMP-xenon spectrum).}

LXe-TPCs provide the capability for ``S2-only'' analyses down to very low thresholds, as pioneered by the XENON collaborations and the DarkSide-50 experiment in liquid argon~\cite{angle:2011th,Aprile:2016wwo,Agnes:2018ves,Agnes:2018fwg,XENON:2024znc,PandaX:2024muv}.  Because LXe-TPCs are sensitive to single electrons,
S2-only analyses can probe energy thresholds of a few hundred eV for nuclear recoils. Given the  increased signal yields expected from H recoils in LXe, S2-only searches could be even more effective in a doped detector. Like other ultra-low-threshold experiments, S2-only searches give up their primary discrimination mechanism in pushing to lower energies. We  calculate the sensitivity of HydroX in an S2-only analysis for two cases where we assume the analysis threshold to be either 3 electrons or 5 electrons.

\begin{figure}[tb!]
  \center
\includegraphics[trim=0 0 0 -0,width=0.99\columnwidth]{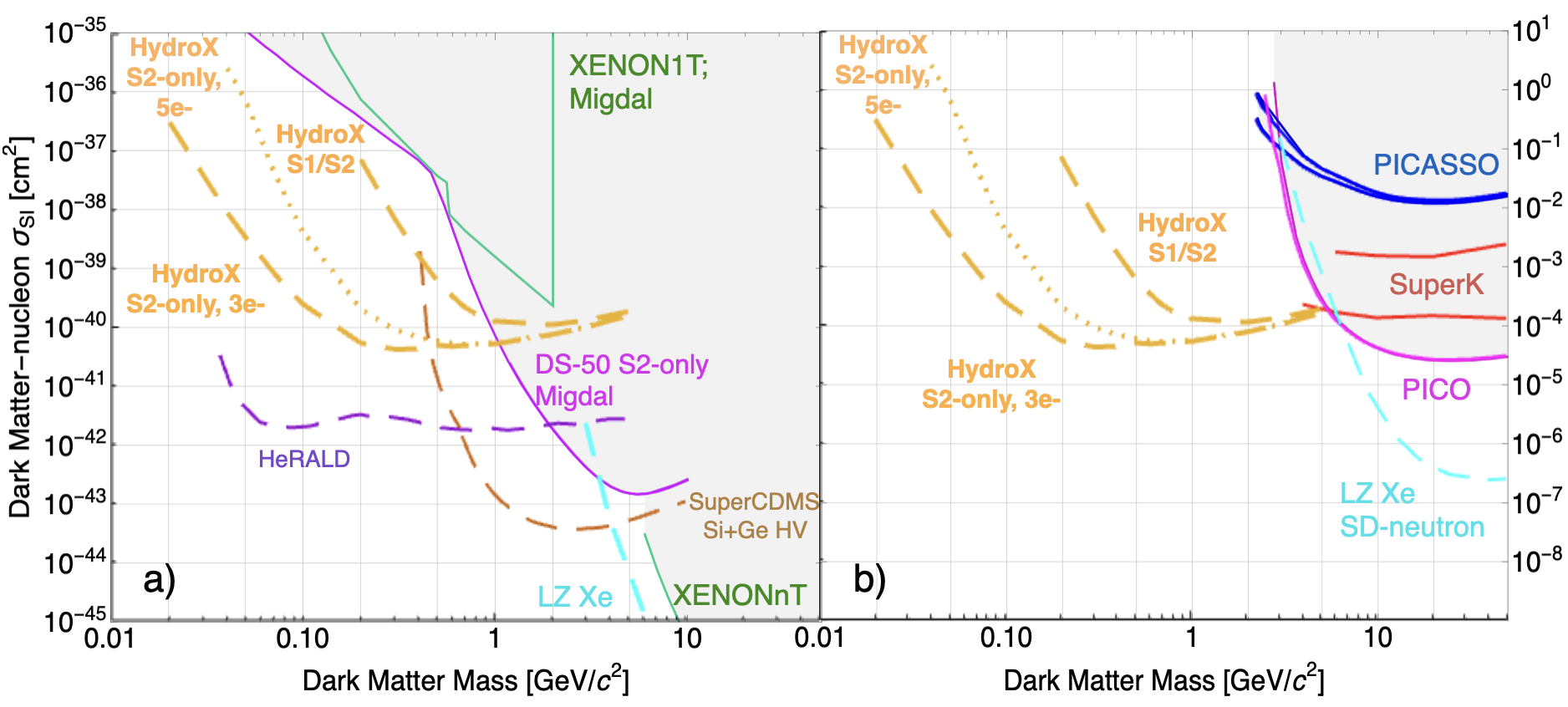}
\caption{\label{fig:Sensitivity} Projected sensitivity of HydroX. The dark matter reach of a $500$ day run of LZ doped with H$_2$ under three different analysis assumptions discussed in the "WIMP Sensitivity" subsection of the Results and Discussion is shown by the orange lines. Panel a) compares the sensitivity with SI projections from other existing and proposed experiments~\cite{LZ:2018qzl,Hertel:2018aal,XENON:2023cxc,XENON:2019zpr,DarkSide:2022dhx,Agnes:2018ves,Agnese:2016cpb}, while panel b) compares to other SD DM-proton interactions, where there are no other sensitive experiments at low DM masses; a recent NEWS-G result is just above the plot, and we also show the LZ SD-neutron sensitivity for comparison~\cite{Amole:2019fdf,Behnke:2016lsk,Choi:2015ara,LZ:2018qzl,NEWS-G:2024jms}. The sensitivity of HydroX per nucleon is the same for SI and SDp interactions, and extends well below $100\,$MeV in mass. }
\end{figure}

The detector response to hydrogen recoils is incorporated into the same background model and Profile Likelihood Ratio analysis used in~\cite{LZ:2018qzl}. Figure~\ref{fig:Sensitivity} shows the projected sensitivity of a 500 day exposure for an S1/S2 analysis and the two S2-only analysis cases. The S1/S2 analysis provides competitive sensitivity down to 200~MeV, while the S2-only analysis has the potential to provide access to masses well below 100~MeV. The right panel of Fig.~\ref{fig:Sensitivity} shows the SD sensitivity of hydrogen in LZ for the same assumptions, providing unique reach to SD interactions of low mass dark matter particles. 

Because low mass dark matter produces an energy spectrum that is sharply rising at the lowest energies, the sensitivity depends critically on the exact details of the detector threshold, which can only be estimated here. Therefore, these projections should be taken as merely indicative of the potential reach of HydroX. \RevA{For a sense of scale, if the S1 signal is a factor of two smaller, the S1/S2 sensitivity decreases by a factor of about 3.5 to 4. If the amount of signal quenching forces us to use a lower loading fraction, the sensitivity would scale approximately linearly with the amount of H$_2$ in the detector, given the same threshold performance.  The overall projected sensitivity is not dependent on the size of the S2 signal (often called ``$g_2$''), as long as single electron resolution is maintained.} 

With the assumptions given, the dominant background is from $^{214}$Pb decays coming from radon emanation, as it is in the higher mass searches already published by LZ. \RevA{As can be seen in Fig.~\ref{fig:S2S1}, there is not a high degree of overlap between $^{214}$Pb decays and a very low mass dark matter signal, so the projections are not very sensitive to the exact level of $^{214}$Pb in the detector. A similar argument would apply to tritium contamination, which would be more of an issue for the sensitivity of any continuing dark matter search with xenon recoils carried out during a HydroX run. However, this analysis does not include ``accidental'' events where spurious S1 and S2-only signals align in time to form a fake signal; along with the exact threshold behavior, this class of event is likely to be the true driver of the ultimate sensitivity. LZ has published results that would observe $\sim$5 accidentals counts in this energy range over a 500-day exposure at a more conservative S2-threshold~\cite{LZ:2022lsv}, which would produce a factor of $\sim$5 reduction in the sensitivity of the S1/S2 analysis. We would likely gain some spectral discrimination against accidentals for very low mass dark matter but a detailed study has not been carried out.}   The XENON and PandaX collaborations have worked to accurately model accidental backgrounds at very low energies in the context of $^8$B neutrino searches~\cite{XENON:2024ijk,PandaX:2024muv}. \RevA{We note that S2-only backgrounds can be highly variable, depending on detector properties such as radiogenic plateout and grid emission. As S2-only backgrounds are not considered in any detail here beyond imposing an S2 threshold, the S2-only sensitivities in particular should be viewed as speculative. }

Figure~\ref{fig:Sensitivity} assumes the signal would be ``ER-like.'' If instead the signals are more ``xenon-recoil-like'' as indicated by~\cite{Haselschwardt:2023iqn}, the increase in S1 would improve sensitivity of the S1/S2 analysis by up to a factor of two in lower mass due to lower thresholds. While the experiment would gain by additional background rejection to ER events, the dominant background would become $^8$B neutrinos scattering with xenon atoms, with significant overlap to WIMP-H scattering in the low energy regime.   In particular, the $^8$B signal would come from interactions with the full active volume of xenon, and it is likely that the absolute sensitivity in cross section would degrade.

Substitution of deuterium for hydrogen, for the same assumptions, would lead to SI sensitivities that are improved by a factor 4 albeit toward higher masses by a factor of $\sqrt{2}$.  Deuterium would allow access to SD DM-neutron interactions, with sensitivities comparable to the right panel of Fig.~\ref{fig:Sensitivity}, also shifted toward higher masses by the factor of $\sqrt{2}$.

\section{Conclusion}\label{sec13}
We describe HydroX, a proposal to add hydrogen to liquid xenon time projection chambers to improve their sensitivity to low mass dark matter particles interacting via both spin independent and spin dependent interactions. Based on assumptions about the potential signal yields for proton recoils in liquid xenon, the amount of hydrogen that can be added to the liquid, and backgrounds, a potential sensitivity is presented with interesting reach, particularly in the spin dependent channel. There are ongoing R\&D efforts to understand several of the technical challenges required to implement HydroX in a real system, \RevA{and the projections presented here should be taken as merely indicative of the potential reach of HydroX. }

\backmatter

\bmhead{Acknowledgements}
Funding for this work is supported by the U.S. Department of Energy, Office of Science, Office of High Energy Physics under Contract Numbers DE-SC0021115, DE-AC02-05CH11231, DE-SC0020216, DE-AC02-07CH11359, DE-SC0015910, DE-SC0011702, DE-SC0015708, DE-SC0008475, DE-SC0013542, DE-AC02-76SF00515, and DE-SC0019066.

\bmhead{Author contributions}
The content of this article was developed in part from the narratives of experiment proposals led by Lippincott and Nelson, involving many of the authors. All authors iterated on the original proposals and approved the final version of the manuscript.

\bmhead{Competing interests}
The authors declare no competing interests.

\bibliography{LZ,Hydrox}

\end{document}